\documentclass[paper]{JHEP}
\usepackage{epsfig}

\def\gsim{\ \rlap{\raise 3pt \hbox{$>$}}{\lower 3pt \hbox{$\sim$}}\ }
\def\lsim{\ \rlap{\raise 3pt \hbox{$<$}}{\lower 3pt \hbox{$\sim$}}\ }
\title{\boldmath On the Inclusive Determination of $|V_{ub}|$ from the 
Lepton Invariant Mass Spectrum
\unboldmath}

\author{Matthias Neubert\\
Newman Laboratory of Nuclear Studies, Cornell University\\
Ithaca, New York 14853, U.S.A.\\
E-mail: \email{neubert@mail.lns.cornell.edu}}

\abstract{Bauer, Ligeti and Luke have recently proposed a new method
for measuring $|V_{ub}|$ in inclusive semileptonic $B$ decays, using a 
cut $\sqrt{q^2}>M_B-M_D$ on the lepton invariant mass to discriminate 
against $b\to c$ transitions. We investigate the structure of the 
heavy-quark expansion for this case and show that to all orders the 
magnitude of the leading perturbative and nonperturbative corrections 
is controlled by a hadronic scale $\mu_c\lsim m_c$ depending on the 
minimal value of $q^2$. These corrections can be analyzed using a 
modified version of the heavy-quark expansion (``hybrid expansion''). 
We find that the theoretical uncertainty in the extraction of 
$|V_{ub}|$ is a factor 2.5 larger than previously estimated, which 
allows for a determination with 10\% accuracy.\\[0.8cm]
\centerline{\it(Submitted to Journal of High Energy Physics)}\\[0.6cm]}


\preprint{CLNS~00/1676\\
June 2000\\[0.15cm]
\hepph{0006068}}

\begin{document}

\section{Introduction}

A precise knowledge of the magnitude of the Cabibbo--Kobayashi--Maskawa 
(CKM) matrix element $V_{ub}$ is of great importance to the study of CP 
violation at the $B$ factories. Whereas the main target of CP-asymmetry
measurements is to fix the angles of the unitarity triangle, the length
of the sides of this triangle are determined by the smallest CKM 
elements, $|V_{ub}|$ and $|V_{td}|$. The measurement of $|V_{ub}|$, 
which neither involves CP violation nor rare loop processes, appears to 
be the simplest step in the overall determination of the unitarity 
triangle. Yet, at present this measurement is limited by uncomfortably 
large theoretical uncertainties. 

$|V_{ub}|$ can be determined most directly from semileptonic $B$ decays 
into charmless final states. The theoretical interpretation of 
exclusive decays such as $B\to\pi\,l\,\nu$ or $B\to\rho\,l\,\nu$ is 
limited by the necessity to predict the $B\to\pi$ or $B\to\rho$
transition form factors, which parameterize the complicated hadronic 
interactions relevant to these decays. The inclusive decays 
$B\to X\,l\,\nu$ admit a cleaner theoretical analysis based on a 
heavy-quark expansion \cite{Chay,Bigi,MaWe}. However, an obstacle is 
that experimentally it is necessary to impose restrictive cuts to 
suppress the background from $B\to X_c\,l\,\nu$ decays (i.e., decays 
into final states with charm). Accounting for such cuts theoretically 
is difficult and usually introduces significant uncertainties.

The first determination of $|V_{ub}|$ from inclusive decays was based 
on a measurement of the charged-lepton energy spectrum close to the 
endpoint region, which is kinematically forbidden for decays with a 
charm hadron in the final state. This restriction eliminates about 90\% 
of the $B\to X_u\,l\,\nu$ signal, and it is difficult to calculate 
reliably the small fraction of the remaining events. The reason 
is that in this portion of phase space the conventional heavy-quark 
expansion breaks down and must be replaced by a twist expansion, in 
which an infinite tower of local operators is resummed into a shape 
function describing the light-cone momentum distribution of the $b$ 
quark inside a $B$ meson \cite{shape1,Russ,shape2}. A promising 
strategy is to infer the fraction of semileptonic decays in the 
endpoint region from a study of the photon spectrum in 
$B\to X_s\,\gamma$ decays, since the leading nonperturbative effects 
in the two decays are described by the same shape function 
\cite{shape2,Korc,Ira,Leib}. Recently, it has been emphasized that a 
cut $M_X<M_D$ on the hadronic invariant mass would provide for a better 
discrimination between $B\to X_u\,l\,\nu$ and $B\to X_c\,l\,\nu$ decays 
\cite{inv1,inv2,inv3}. Such a cut eliminates the charm background, 
while affecting only about 20\% of the signal events. Despite of this 
advantage, however, it is difficult to calculate precisely what this 
fraction is \cite{Fulvia}. 

Bauer, Ligeti and Luke (BLL) have pointed out that the situation may be 
better if a discrimination based on a cut on the lepton invariant mass 
$\sqrt{q^2}$ is employed \cite{BLL}. Requiring $q^2>(M_B-M_D)^2$ 
eliminates the charm background, while containing about 20\% of the 
signal events. This fraction is much less than in the case of a cut on 
hadronic invariant mass, but the inclusive rate with a lepton invariant 
mass cut offers the advantage of being calculable using a conventional 
heavy-quark expansion, without a need to resum an infinite series of 
nonperturbative corrections. BLL find that the fraction of 
$B\to X_u\,l\,\nu$ events with $q^2>q_0^2$, where $q_0^2\ge(M_B-M_D)^2$ 
is required to eliminate the charm background, is given by
\begin{equation}\label{BLL}
   F(q_0^2) = (1+\hat q_0^2)(1-\hat q_0^2)^3 
   + \tilde X(\hat q_0^2)\,\frac{\alpha_s(m_b)}{\pi} 
   - \left( \hat q_0^2 - 2\hat q_0^6 + \hat q_0^8 \right)
   \frac{12\lambda_2}{m_b^2} + \dots \,,
\end{equation}
where $\hat q_0^2=q_0^2/m_b^2$, and the dots represent corrections of 
higher order in the expansion in powers of $\alpha_s(m_b)$ and 
$\Lambda/m_b$ (with $\Lambda$ a characteristic hadronic scale). The 
function $\tilde X(\hat q_0^2)$ can be obtained from results presented 
in \cite{Kuhn}. The nonperturbative parameter $\lambda_2
=\frac14(M_{B^*}^2-M_B^2)\simeq 0.12$\,GeV$^2$ was introduced in 
\cite{FaNe}. As it stands, eq.~(\ref{BLL}) seems to provide a solid 
theoretical basis for a systematic analysis of the partial decay rate 
in an expansion in logarithms and powers of $\Lambda/m_b$. (This leaves 
aside the fact that the lepton invariant mass cut eliminates about 80\% 
of all $B\to X_u\,l\,\nu$ events, and therefore one may worry that 
violations of quark--hadron duality may be larger than for the total 
inclusive semileptonic rate.) 

The purpose of this work is to analyze the structure of the result 
(\ref{BLL}) in more detail. We show that the relevant mass scale 
$\mu_c$ controlling the size of corrections in the heavy-quark 
expansion is less than or of order the charm-quark mass, rather than 
the heavier $b$-quark mass. This is so because the largest values of 
the hadronic invariant mass and energy accessible are of order the 
charm mass or less. Although the ratio $m_c/m_b$ is usually taken as a 
constant when discussing the heavy-quark limit, it is well known that 
the convergence of heavy-quark expansions at the charm scale can be 
poor. This makes our observation relevant. We suggest that an 
appropriate framework in which to investigate the leading corrections 
to the heavy-quark limit in the present case is a modified version of 
the heavy-quark expansion to which we refer as ``hybrid expansion''. 
The idea is that in the kinematic region where 
$M_B\gg E_X\sim M_X\gg\Lambda$ one can perform a two-step expansion in 
the ratios $E_X/M_B$ and $\Lambda/E_X$. A similar strategy has been 
used in applications of the heavy-quark effective theory (HQET) to 
resum the so-called ``hybrid logarithms'' $\alpha_s\ln(m_c/m_b)$ 
\cite{VoSh} arising in current-induced $b\to c$ transitions using 
renormalization-group (RG) equations \cite{FaGr85,Ne92}. In the context 
of inclusive decays, an approach similar in spirit to our proposal was 
suggested first by Mannel \cite{Thom}. We use the hybrid expansion to 
obtain a RG-improved expression for the perturbative corrections to the 
quantity $F(q_0^2)$ at next-to-leading order (NLO), as well as to 
estimate the size of higher-order power corrections omitted in 
(\ref{BLL}).

\section{Structure of the hadronic tensor}
\label{sec:tensor}

The strong-interaction dynamics relevant to inclusive $B$ decays is 
encoded in a hadronic tensor defined as the forward matrix element of 
the time-ordered product of two weak currents between $B$-meson 
states. Although the variable of prime interest to our discussion is 
the lepton invariant mass, the physics of the hadronic tensor is most 
naturally described in terms of the hadronic invariant mass and 
energy, $M_X$ and $E_X$. These variables are related by 
$q^2=M_B^2-2 M_B E_X+M_X^2$, and thus the restriction $q^2>q_0^2$
implies
\begin{equation}
   M_X \le E_X \le \frac{M_B^2 + M_X^2 - q_0^2}{2 M_B} \,, \qquad
   M_\pi^2 \le M_X^2 \le (M_B - \sqrt{q_0^2})^2 \,.
\end{equation}
With the optimal choice $q_0^2=(M_B-M_D)^2$ this gives 
$M_\pi\le M_X\le M_D$ and $M_X\le E_X\le M_D-\frac12(M_D^2-M_X^2)/M_B$. 
Both variables vary between $M_\pi$ and $M_D$. If the cutoff $q_0^2$ is 
chosen higher, as may be required for experimental reasons, their 
maximal values become less than $M_D$.

The corresponding variables entering a partonic description of inclusive
decay rates are the parton invariant mass and energy, $\sqrt{p^2}$ and 
$v\cdot p$, where $p$ is related to the lepton momentum $q$ by
$p=m_b v-q$, and $v$ is the velocity of the $B$ meson. The phase space 
for the dimensionless variables $\hat p^2=p^2/m_b^2$ and 
$z=2v\cdot p/m_b$ is
\begin{equation}\label{range}
   2\sqrt{\hat p^2} \le z \le 1 - \hat q_0^2 + \hat p^2 \,, \qquad
   0 \le \hat p^2 \le (1 - \sqrt{\hat q_0^2})^2 \,.
\end{equation}
For the purpose of our discussion here $m_b$ is the pole mass of the 
$b$ quark. Alternative mass definitions will be discussed in more
detail later. The optimal value of $\hat q_0^2$ is $\hat q_0^2
=(M_B-M_D)^2/m_b^2\simeq(1-\frac{m_c}{m_b})^2$. In this case the 
largest values of the parton variables are 
$z_{\rm max}=2\,\frac{m_c}{m_b}$ and 
$\hat p^2_{\rm max}=(\frac{m_c}{m_b})^2$. Without a restriction on 
$q^2$, the phase space for these variables would be such that 
$z_{\rm max}=2$ and $\hat p^2_{\rm max}=1$. In other words, the lepton 
invariant mass cut restricts both variables to a region where they are 
parametrically suppressed, such that $v\cdot p$ and $\sqrt{p^2}$ are at 
most of order $m_c$. Besides $m_b$, the charm-quark mass is then a 
relevant scale that determines the magnitude of strong-interaction 
effects in the heavy-quark expansion. 

To make the parametric suppression noted above more explicit, we
introduce a characteristic scale $\mu_c$ and an associated small 
expansion parameter $\epsilon$ by
\begin{equation}
   \mu_c = \frac{m_b^2-q_0^2}{2m_b} = O(m_c) \,,
   \qquad \epsilon = \frac{1-\hat q_0^2}{2} = \frac{\mu_c}{m_b} \,,
\end{equation}
and rewrite the fraction $F(q_0^2)$ of $B\to X_u\,l\,\nu$ events with 
$q^2>q_0^2$ as
\begin{equation}\label{Feps}
   F(\epsilon) = 16\epsilon^3(1-\epsilon) \left[ C_0(\epsilon)
   - \frac32\,C_2(\epsilon)\,\frac{\lambda_2}{\mu_c^2}
   + O[(\Lambda/\mu_c)^3] \right] \,,
\end{equation}
where $C_n(\epsilon)=1+O(\epsilon,\alpha_s)$ are short-distance 
coefficients. From (\ref{BLL}) we find
\begin{eqnarray}\label{Cn}
   C_0(\epsilon) &=& 1 + \frac{\alpha_s}{3\pi} \left[
    \left( -6\ln2\epsilon - \frac32 - \frac{2\pi^2}{3} \right) 
    + \epsilon \left( -8\ln2\epsilon + \frac{73}{6} \right)
    + O(\epsilon^2) \right] + O(\alpha_s^2) \,, \nonumber\\
   C_2(\epsilon) &=& \frac{1-8\epsilon^2+8\epsilon^3}{1-\epsilon}
    + O(\alpha_s) \,.
\end{eqnarray}
The definition of $\epsilon$ and $\mu_c$, as well as the explicit form 
of the coefficients $C_n(\epsilon)$, depend on the definition of the 
heavy-quark mass. The above result for $C_0(\epsilon)$ refers to the 
pole mass. Later we will introduce a more suitable mass definition and 
give an exact NLO expression for $C_0(\epsilon)$ including the 
higher-order terms in $\epsilon$.

\FIGURE[t]{\epsfig{file=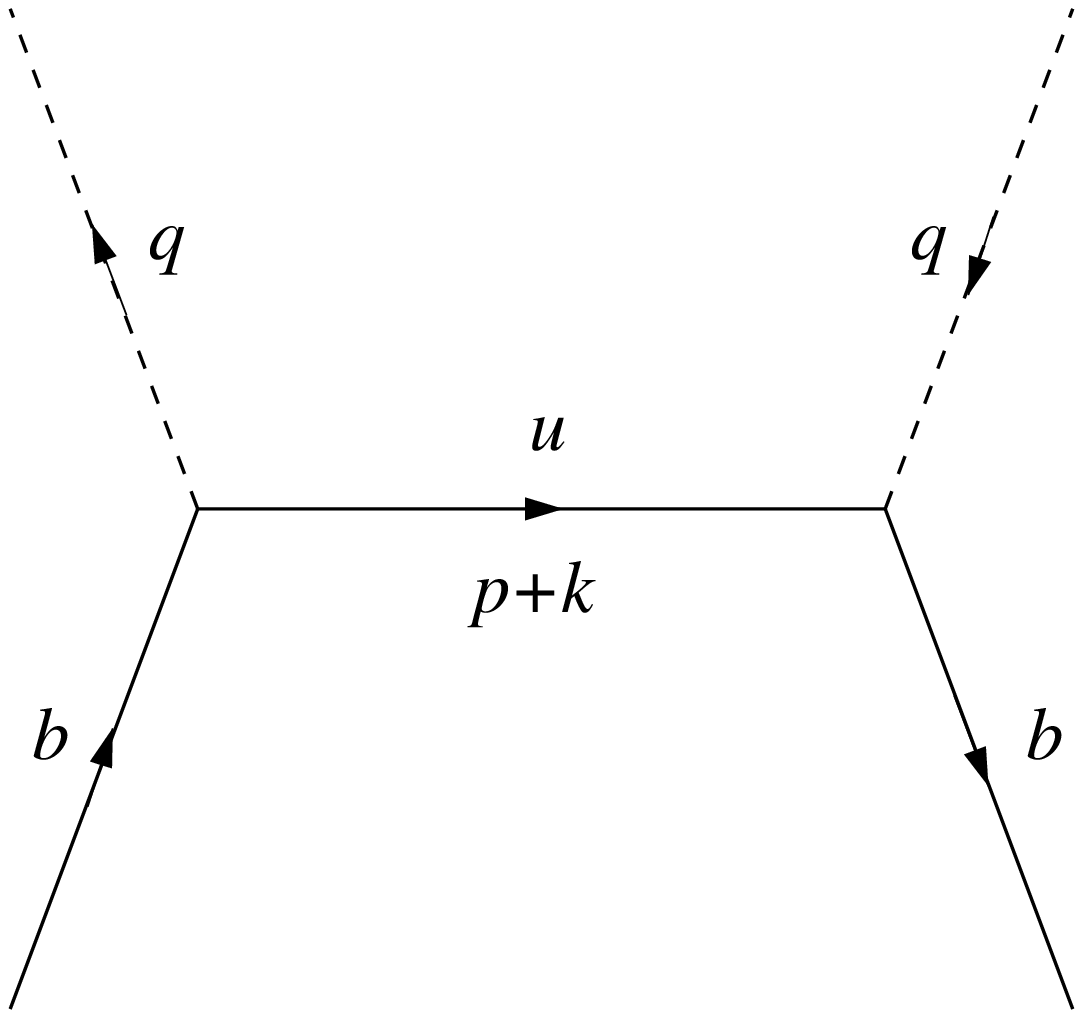,width=6.0cm}
\vspace{-0.3cm}
\caption{\label{fig:1}
Tree-level contribution to the hadronic tensor. Dashed lines represent 
the weak currents.}}

It is apparent from (\ref{Feps}) that the leading power correction 
proportional to $\lambda_2$ is of order $(\Lambda/\mu_c)^2$. It is not
difficult to see that, in the presence of a lepton invariant mass cut, 
also in higher orders the power corrections scale like 
$(\Lambda/\mu_c)^n$. For simplicity of the argument we work to leading 
order in $\alpha_s$, where only the tree diagram shown in 
Figure~\ref{fig:1} contributes to the hadronic tensor. In momentum 
space, the $u$-quark propagator gives a contribution
\begin{equation}
   \frac{(p+k)^\mu}{p^2+2p\cdot k+k^2} \,,
\end{equation}
where $k=O(\Lambda)$ is the residual momentum of the heavy quark 
inside the $B$ meson \cite{review}. Roughly speaking, the heavy-quark 
expansion is obtained by replacing the residual momentum with a 
covariant derivative, $k^\mu\to iD^\mu$, thereby introducing the 
soft interactions of the $u$-quark jet with the background field of
the light degrees of freedom in the $B$ meson. Let us discuss how the 
three terms in the denominator of the propagator scale in the different 
kinematic regions relevant to the determination of $|V_{ub}|$. Whereas 
the $k^2\sim\Lambda^2$ term is always suppressed, the relative 
magnitude of the two other terms depends on the kinematic region 
considered. For the regions of the large charged-lepton energy or low 
hadronic invariant mass, the terms $p^2\sim\max[\Lambda\,m_b,m_c^2]$ 
and $p\cdot k\sim\Lambda\,m_b$ are of the same magnitude, and it is 
thus necessary to resum terms of the form $(p\cdot k/p^2)^n$ to all 
orders in the heavy-quark expansion. This leads to a twist expansion, 
where these terms are absorbed into a nonperturbative shape function 
\cite{shape1,Russ,shape2}. In contrast, for large lepton invariant 
mass, $p\cdot k\sim\Lambda\,m_c$ is parametrically suppressed with 
respect to $p^2\sim m_c^2$ by a power of $\Lambda/m_c$.

These general observations can also be derived from the explicit 
expression for the differential decay rate expressed in terms of the 
variables $z$ and $\hat p^2$, normalized to the total decay rate. At 
NLO in the heavy-quark expansion we find
\begin{eqnarray}\label{rate}
   \frac{1}{\Gamma}\,
   \frac{\mbox{d}^2\Gamma}{\mbox{d}z\,\mbox{d}\hat p^2}
   &=& 2 z^2 (3-2z)\,\delta(\hat p^2)
    + \frac{\alpha_s}{3\pi}\,E(z,\hat p^2) \nonumber\\
   &&\mbox{}- \delta(\hat p^2) \left[
    \frac{z}{3}\,(36-27z-16z^2)\,\frac{\lambda_1}{m_b^2}
    + (12+12z-63z^2+8z^3)\,\frac{\lambda_2}{m_b^2} \right] \nonumber\\
   &&\mbox{}- \delta'(\hat p^2) \left[
    \frac{z^2}{3}\,(18+3z-14z^2)\,\frac{\lambda_1}{m_b^2}
    + z^2 (6+3z-10z^2)\,\frac{\lambda_2}{m_b^2} \right] \nonumber\\
   &&\mbox{}- \delta''(\hat p^2)\,\frac{z^4}{3}\,(3-2z)\,
    \frac{\lambda_1}{m_b^2} + \dots \,,
\end{eqnarray}
where $\lambda_1$ is a nonperturbative parameter related to the average
kinetic energy of the $b$ quark inside the $B$ meson \cite{FaNe}, and 
the function $E(z,\hat p^2)$ gives the perturbative correction 
calculated in \cite{Fulvia}, which also includes the correction to the
total decay rate appearing in the denominator on the left-hand side. 
The power corrections in (\ref{rate}) have been derived using the 
results of \cite{MaWe,FLNN}. In the kinematic region where 
$z=O(\epsilon)$ and $\hat p^2=O(\epsilon^2)$, it is instructive to 
change variables from $(z,\hat p^2)$ to $(z,\xi)$, where 
$\xi=4\hat p^2/z^2=p^2/(v\cdot p)^2\in[0,1]$ is related to the parton 
velocity in the $B$ rest frame. The kinematic range for these variables 
is
\begin{equation}
   0 \le \xi \le 1 \,, \qquad
   0 \le z \le \frac{2}{\xi}\,(1-\sqrt{1-2\epsilon\xi})
   = 2\epsilon + \epsilon^2\xi + O(\epsilon^3) \,.
\end{equation}
The variable $\xi$ is of order unity irrespective of the lepton 
invariant mass cut. Thus, after the transformation the only 
parametrically small quantity is $z=O(\epsilon)$. In terms of the new 
variables, the double-differential decay rate turns into an expansion 
in powers of $\Lambda/(z m_b)$, and the perturbative corrections contain 
single logarithms of $z$. Integrating the double-differential rate over 
$z$, and keeping only terms of leading order in $\epsilon$, we obtain
\begin{eqnarray}\label{xirate}
   \frac{1}{\Gamma}\,
   \frac{\mbox{d}\Gamma}{\mbox{d}\xi}
   &=& 16\epsilon^3 \Bigg\{ \delta(\xi) \left[ 1 
    - \frac{\alpha_s}{3\pi} \left( 6\ln\frac{\mu_c}{m_b}
    + 8\ln 2(\ln 2-1)
    + \frac{31}{2} \right) \right] - \nonumber\\
   &&\mbox{}- \frac{\alpha_s}{3\pi} \left[ \frac{4\ln\xi}{\xi}
    + \frac{1}{\xi} \left( 7\sqrt{1-\xi} - 8\ln(1+\sqrt{1-\xi}) \right)
    \right]_+ \nonumber\\
   &&\mbox{}- \frac{3\lambda_2}{2\mu_c^2}\,\delta(\xi)
    - \frac{3(\lambda_1+\lambda_2)}{\mu_c^2}\,\delta'(\xi)
    - \frac{2\lambda_1}{\mu_c^2}\,\delta''(\xi) + \dots \Bigg\} 
    + O(\epsilon^4) \,.
\end{eqnarray}
The integral of this expression over $\xi$ reproduces the leading terms 
in $\epsilon$ in (\ref{Feps}). The result for the power corrections 
shows that indeed $\mu_c=O(m_c)$ is the characteristic scale of the 
hybrid expansion. Perturbative logarithms of $\mu_c/m_b$ appear because 
the intrinsic scale of the hadronic tensor is smaller than the mass of 
the external $b$ quarks, suggesting that the appropriate scale to 
evaluate the running coupling $\alpha_s$ is significantly less than 
$m_b$. This is in accordance with the observation \cite{LSW} that the 
physical scale derived using the Brodsky--Lepage--Mackenzie (BLM) 
scale-setting prescription \cite{BLM} strongly decreases with 
increasing $q^2$. We will come back to the question of scale setting in 
the next section.

\FIGURE[h]{\epsfig{file=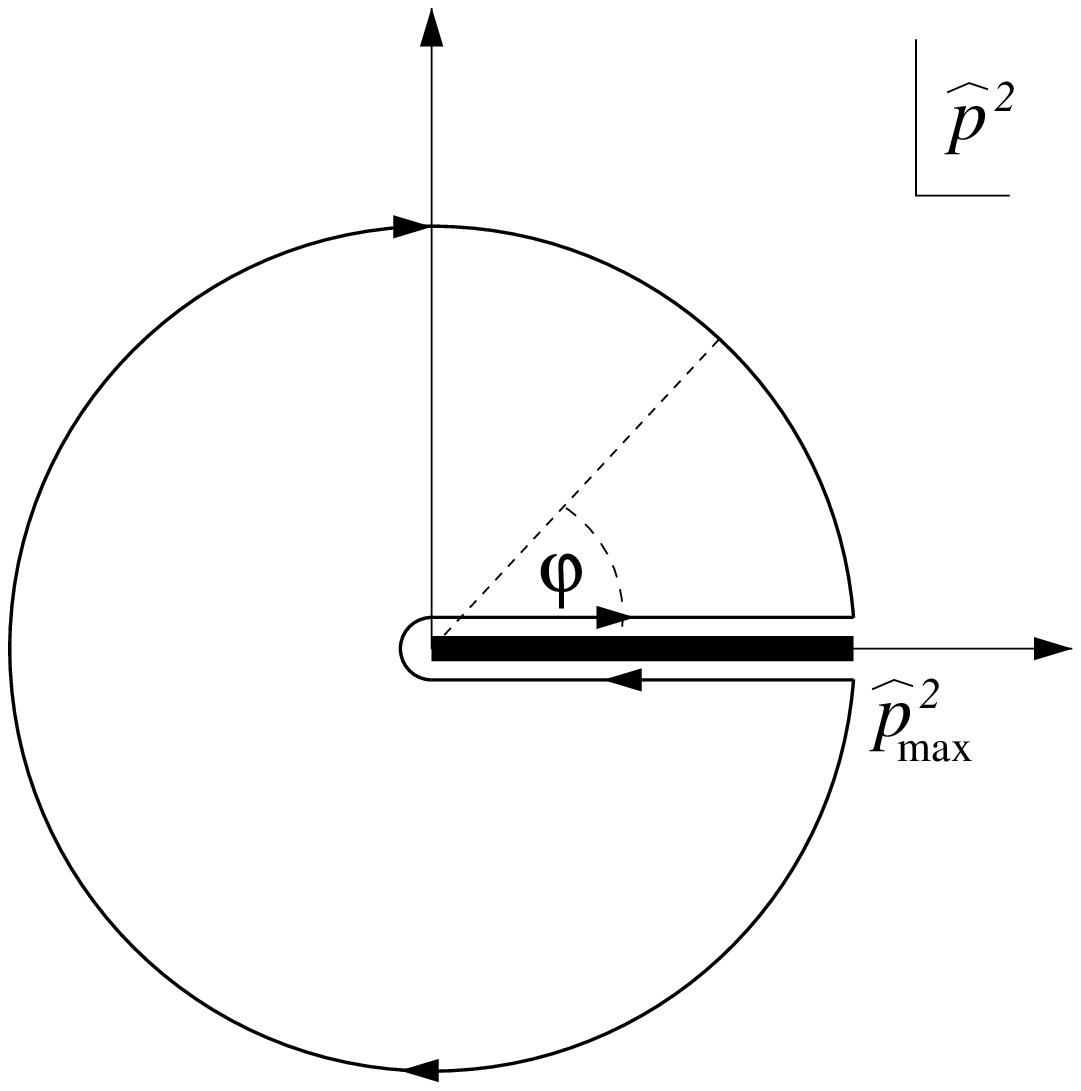,width=6.0cm}
\caption{\label{fig:2}
Contour representation of $F(q_0^2)$ in the complex $\hat p^2$ plane.}}

For later purposes, it will be useful to have yet another way of 
reproducing the result (\ref{Feps}). To this end, we represent the 
fraction $F(q_0^2)$ as a contour integral in the complex $\hat p^2$ 
plane. Such a representation exists because the hadronic tensor is an 
analytic function in the complex plane apart from discontinuities 
located on the positive real $\hat p^2$ axis, as illustrated in 
Figure~\ref{fig:2}. We write
\begin{equation}
   F(q_0^2) = \frac{i}{2\pi}
   \oint\limits_{|\hat p^2|=\hat p^2_{\rm max}}\!\!\!
   \mbox{d}\hat p^2\,T(\hat p^2,\hat q_0^2) \,,
\end{equation}
where $\hat p_{\rm max}^2=(1-\sqrt{\hat q_0^2})^2$, and the correlator 
$T(\hat p^2,\hat q_0^2)$ can be obtained using dispersion relations 
\cite{Fulvia}. Eliminating $\hat q_0^2$ in favor of $\epsilon$, and 
parameterizing $\hat p^2=\hat p^2_{\rm max}\,e^{i\varphi}$ on the 
contour of integration, the result can be written in the form
\begin{equation}\label{contour}
   F(\epsilon) = 16\epsilon^3(1-\epsilon) \int\limits_0^{2\pi} 
   \frac{\mbox{d}\varphi}{2\pi}\,t(e^{i\varphi},\epsilon) \,,
\end{equation}
where the function $t(e^{i\varphi},\epsilon)$ contains the corrections 
to the heavy-quark limit. We obtain
\begin{eqnarray}\label{tfun}
   t(e^{i\varphi},\epsilon)
   &=& C_0(\epsilon) + \frac{\alpha_s}{3\pi}\,Y(e^{i\varphi},\epsilon)
    - \frac32\,C_2(\epsilon)\,\frac{\lambda_2}{\mu_c^2}
    + \left[ D_1(\epsilon)\,\frac{\lambda_1}{\mu_c^2}
    + D_2(\epsilon)\,\frac{\lambda_2}{\mu_c^2} \right] e^{-i\varphi}
    \nonumber\\
   &&\mbox{}- \frac45\,E_1(\epsilon)\,\frac{\lambda_1}{\mu_c^2}\,
    e^{-2i\varphi} + O[(\Lambda/\mu_c)^3] \,.
\end{eqnarray}
The function $Y(e^{i\varphi},\epsilon)$ with
\begin{equation}\label{Yfun}
   Y(x,\epsilon) = \frac{10}{3}\,x - \frac{184}{9}
   + \left( \frac{29}{3} - \frac53\,x \right)
   u \ln\frac{u+1}{u-1} - 2 \ln^2\frac{u+1}{u-1} + O (\epsilon) 
\end{equation}
and $u=\sqrt{1-x}$ is defined such that its contour integral vanishes. 
The coefficients $D_i(\epsilon)$ and $E_1(\epsilon)$ are equal to 1 in 
the limit $\epsilon\to 0$. Their explicit expressions are
\begin{eqnarray}
   D_1(\epsilon)
   &=& \frac{\epsilon^2}{(1-\epsilon)(1-\sqrt{1-2\epsilon})^2}
    \left( 1 - \frac74\,\epsilon + \frac45\,\epsilon^2 \right)
    + O(\alpha_s) \,, \nonumber\\
   D_2(\epsilon)
   &=& \frac{\epsilon^2}{(1-\epsilon)(1-\sqrt{1-2\epsilon})^2}
    \left( 1 + \frac34\,\epsilon - 4\epsilon^2 \right) + O(\alpha_s)
    \,, \nonumber\\
   E_1(\epsilon)
   &=& \frac{\epsilon^4}{(1-\epsilon)(1-\sqrt{1-2\epsilon})^4}
    \left( 1 - \frac{10}{9}\,\epsilon \right) + O(\alpha_s) \,.
\end{eqnarray}

Equations (\ref{xirate}) and (\ref{tfun}) allow us to study the 
behavior of the leading corrections in the heavy-quark expansion in 
more detail. In particular, we will utilize them to estimate unknown, 
higher-order power corrections. We will, however, first investigate 
the perturbative corrections in more detail, using the hybrid expansion 
as a tool to perform a systematic RG improvement of the one-loop 
expressions in (\ref{BLL}) and (\ref{Cn}).

\section{RG improvement and definition of the heavy-quark mass}

Based on the observation that the event fraction $F(q_0^2)$ in 
(\ref{BLL}) receives very small corrections of order 
$\beta_0\alpha_s^2$, where $\beta_0$ is the first coefficient of the 
$\beta$-function, BLL have argued that the perturbative uncertainty in 
their prediction is negligible \cite{BLL}. The purpose of this section 
is to critically reanalyze the perturbative uncertainty in the 
calculation of this quantity. We first note that in the present case it 
is misleading to associate the size of $\beta_0\alpha_s^2$ corrections 
with a physical BLM scale in the process. The total semileptonic rate 
and the lepton invariant mass spectrum receive very large corrections 
of order $\beta_0\alpha_s^2$, corresponding to very low BLM scales 
\cite{LSW}. The fraction $F(\epsilon)$ is defined as the ratio of the 
partially integrated lepton spectrum and the total rate. If both of 
these quantities have low physical scales, the same must be true for 
their ratio. In our opinion the small $\beta_0\alpha_s^2$ correction 
observed in \cite{BLL} is thus due to an accidental cancellation and 
does not bare any physical significance. 

Here we follow a different strategy to estimate the potential 
importance of higher-order effects. We have argued that a useful 
framework in which to analyze the fraction $F(\epsilon)$ is provided 
by a hybrid expansion, in which the physics associated with the three
mass scales $m_b\gg\mu_c\gg\Lambda$ is disentangled. Since the 
intermediate scale $\mu_c$ is only about 1\,GeV or less, and since the 
running of the strong coupling in the region between $m_b$ and 1\,GeV 
is significant, we expect important higher-order perturbative 
corrections resolving the scale ambiguity. At one-loop order, this is 
indicated by the presence of logarithms of $\epsilon$ in (\ref{Cn}). 
At any given order in an expansion in powers of $\epsilon$ the 
contributions associated with the two couplings $\alpha_s(m_b)$ and 
$\alpha_s(\mu_c)$ can be separated by solving RG equations in the 
hybrid expansion at NLO. The residual scale ambiguity left after RG 
improvement provides an estimate of the perturbative uncertainty in 
the result. Unfortunately, to perform this program one must compute, 
at every order in $\epsilon$, the two-loop anomalous dimensions of a 
new tower of higher-dimensional operators. At present, these anomalous 
dimensions are known only for the operators entering at the leading 
order in $\epsilon$, although partial results exist for the operators 
relevant to the $O(\epsilon)$ terms \cite{Amor}. 

We now discuss the RG improvement at leading order in $\epsilon$ in 
detail. The first step in the construction of the hybrid expansion is 
to expand the weak currents in the definition of the hadronic tensor 
in terms of operators of the HQET, with the result 
\cite{review}
\begin{equation}\label{Jexp}
   \bar q\gamma_\mu(1-\gamma_5)\,b
   \to C_1\Big(\frac{m_b}{\mu}\Big)\,\bar q\gamma_\mu(1-\gamma_5)\,h_v
   + C_2\Big(\frac{m_b}{\mu}\Big)\,\bar q\,v_\mu(1+\gamma_5)\,h_v
   + O(1/m_b) \,,
\end{equation}
where $v_\mu$ is the $B$-meson velocity, $h_v$ are the 
velocity-dependent fields of the HQET, and $\mu$ is the scale at which 
the operators are renormalized. The RG-improved expressions for the 
Wilson coefficients $C_i(m_b/\mu)$ are known at NLO. The terms of order 
$1/m_b$ in (\ref{Jexp}) would contribute at order $\epsilon$ in the 
hybrid expansion and can be neglected for the discussion of the leading 
terms. It is important that all dependence on the $b$-quark mass is 
explicit in (\ref{Jexp}). We now insert this result into the 
time-ordered product of currents in the hadronic tensor and perform an 
operator product expansion of the current product. This is an expansion 
in logarithms and powers of $\Lambda/\mu_c$. The scale $m_b$ does not 
appear in the matrix elements of the hybrid expansion. We find that at 
leading order in $\epsilon$ only the HQET current product proportional 
to $C_1^2(m_b/\mu)$ contributes. Using the known NLO expression for 
the coefficient $C_1(m_b/\mu)$ \cite{XiMu,BrGr}, we obtain
\begin{eqnarray}\label{C0pole}
   C_0(\epsilon) &=& \left( \frac{\alpha_s(\mu)}{\alpha_s(m_b)} 
    \right)^\frac{4}{\beta_0} \Bigg[
    1 + \frac{2\alpha_s(m_b)}{\pi} \left( Z_{\rm hl} 
    - \frac{25}{12} + \frac{\pi^2}{3} \right) \nonumber\\
   &&\hspace{2.77cm}\mbox{}- \frac{2\alpha_s(\mu)}{\pi}
    \left( Z_{\rm hl} - \ln\frac{\mu}{2\mu_c} - \frac{11}{6}
    + \frac{4\pi^2}{9} \right) \Bigg] + O(\epsilon) \nonumber\\
   &\simeq& \left( \frac{\alpha_s(\mu)}{\alpha_s(m_b)}
    \right)^\frac{12}{25} \left[ 1 - 0.71\,\frac{\alpha_s(m_b)}{\pi} 
    - \frac{\alpha_s(\mu)}{\pi} \left( 3.37 
    - 2\ln\frac{\mu}{\mu_c} \right) \right] + O(\epsilon) \,. \qquad
\end{eqnarray}
Here $\beta_0=\frac{25}{3}$ and $Z_{\rm hl}=-\frac{9403}{7500}
-\frac{7\pi^2}{225}$ are perturbative coefficients evaluated for
$n_f=4$ light quark flavors, as is appropriate for a scale of order
$m_c$. Note that the NLO corrections proportional to the coupling 
$\alpha_s(m_b)$ contain the corrections to the total semileptonic rate. 
The above result for $C_0(\epsilon)$ gives the RG-improved form of the 
leading term in $\epsilon$ in the one-loop expression in (\ref{Cn}). 
This result is formally scale independent at NLO. The renormalization 
scale $\mu$ should be chosen of order $\mu_c$ in order to avoid large 
logarithms in the hybrid expansion. (The inappropriate choice $\mu=m_b$ 
would reproduce the one-loop result with $\alpha_s$ evaluated at $m_b$.) 

\FIGURE[h]{\epsfig{file=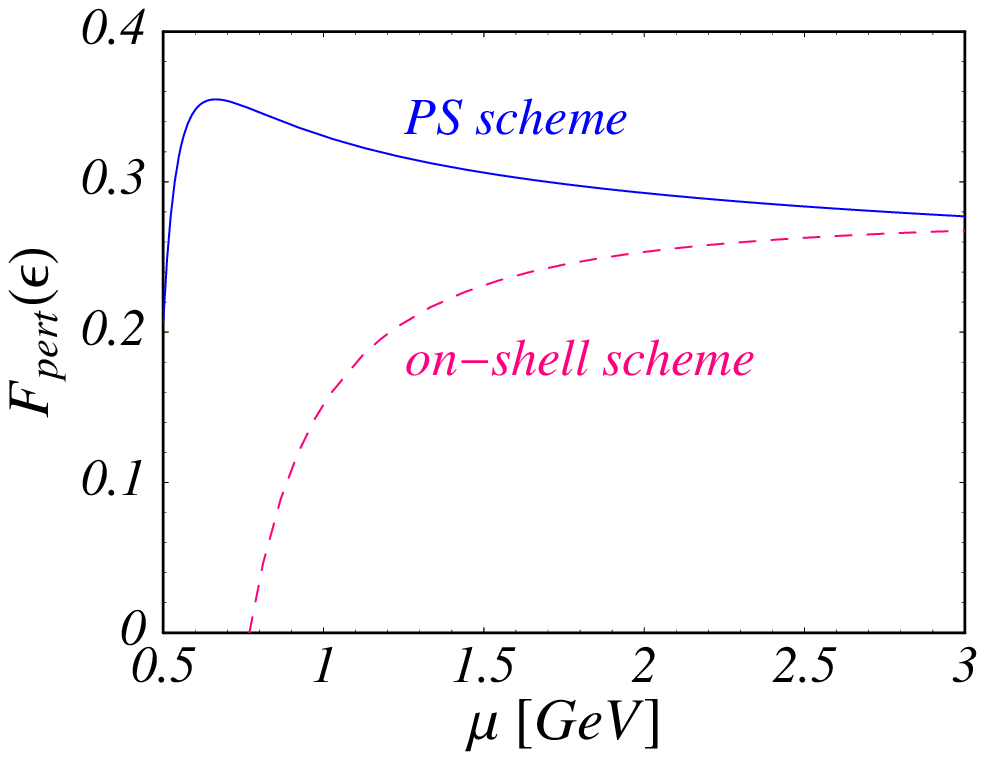,width=7.1cm}
\caption{\label{fig:4}
Scale dependence of the perturbative result for the fraction 
$F(\epsilon)$ to leading order in the hybrid expansion, evaluated for 
$q_0^2=(M_B-M_D)^2$. The two curves refer to different mass 
renormalization schemes.}}

The dashed line in Figure~\ref{fig:4} shows the perturbative prediction 
for the fraction $F(\epsilon)$ at leading order in $\epsilon$, as a 
function of the renormalization scale. Here and below we use the 
two-loop expression for the running coupling constant normalized such 
that $\alpha_s(M_Z)=0.118$. In accordance with relation (\ref{PSdef}) 
below we use $m_b=5.0$\,GeV for the pole mass, noting that this value 
(and thus the normalization of the dashed curve) has a large 
uncertainty. Two observations are important. First, the scale dependence 
of the dashed curve is significant, and hence the result obtained with 
an appropriate choice of scale $\mu\approx\mu_c$ is much lower than that 
obtained with the naive choice $\mu=m_b$. Secondly, perturbation theory 
in the on-shell scheme breaks down at a scale not much less than the 
appropriate scale $\mu\approx\mu_c\approx 1$\,GeV. We conclude that in 
the on-shell scheme there is a large perturbative uncertainty in the 
calculation of the coefficient $C_0(\epsilon)$, which is not apparent 
from the naive one-loop result. This conclusion is in contrast with the 
assumption made by BLL, that the perturbative uncertainty is negligible 
\cite{BLL}.

The breakdown of perturbation theory at a scale of order $\mu_c$ can
be traced back to the large coefficient of the NLO correction 
proportional to $\alpha_s(\mu)$ in (\ref{C0pole}). We will now show that 
the size of this coefficient can be reduced significantly by adopting a 
more appropriate definition of the heavy-quark mass. So far we have 
worked in the on-shell scheme, where the mass is defined as the pole in 
the renormalized quark propagator, $m_b=m_b^{\rm pole}$. Since the pole 
mass is affected by IR renormalon ambiguities \cite{BeBr,BSUV}, it is 
better to eliminate it from the final expressions for inclusive decay 
rates. If a new mass definition $m_b'$ is introduced via 
$m_b^{\rm pole}=Z_m\,m_b'$, it follows from (\ref{Feps}) that
\begin{equation}\label{C0rela}
   C_0'(\epsilon') = C_0^{\rm pole}(\epsilon') \left[ 1
   + \frac{Z_m-1}{\epsilon'}\,
   \frac{(1-2\epsilon')(3-4\epsilon')}{1-\epsilon'} + \dots \right] \,,
\end{equation}
where the prime on $\epsilon$ indicates that this parameter in sensitive
to the definition of $m_b$. We observe that a multiplicative 
redefinition of $m_b$ with $Z_m(\alpha_s)=1+O(\alpha_s)$, such as the
relation between the pole mass and the running mass defined in the 
$\overline{\rm MS}$ scheme, is not appropriate in our case, since it 
would lead to a contribution to $C_0'(\epsilon')$ that is enhanced by a 
factor of $\alpha_s/\epsilon'$. This would upset the power counting in
the hybrid expansion. We suggest instead to work with a short-distance 
mass subtracted at a scale $\mu_f$ of order $\mu_c$, which is the 
natural scale of our problem. 

It is well known that the convergence of the perturbative series for 
near on-shell problems in heavy-quark physics can be largely improved 
by introducing low-scale subtracted quark masses, which have the 
generic property that they differ from the pole mass by an amount 
proportional to the subtraction scale $\mu_f$. Several such mass 
definitions exist and have been applied to various processes 
\cite{PSmass,1Smass,kinmass,Martin}. To illustrate the point, we use
the potential-subtracted (PS) mass $m_b^{\rm PS}(\mu_f)$ introduced by
Beneke \cite{PSmass} and evaluate it at the scale 
$\mu_f=\mu_c^{\rm PS}$.\footnote{We could instead evaluate the PS mass
at a scale $\mu_f=\xi\,\mu_c$ with $\xi=O(1)$, however this would lead
to more complicated expressions. Varying $\xi$ between 1 and 2 leads 
to a variation of the results by an amount similar to the perturbative 
uncertainty estimated later in this section.} 
At NLO, the relation between the pole mass and the PS mass reads
\begin{equation}\label{PSdef}
   m_b^{\rm pole} = m_b^{\rm PS}(\mu_c^{\rm PS})  
   + \mu_c^{\rm PS}\,\frac{4\alpha_s(\mu)}{3\pi} \left\{ 1
   + \frac{\alpha_s(\mu)}{2\pi} \left[
   \beta_0 \left( \ln\frac{\mu}{\mu_c^{\rm PS}} + \frac{11}{6}
   \right) - 4 \right] + \dots \right\} \,,
\end{equation}
which is formally independent of the scale $\mu$ at which the coupling 
is renormalized. Note that the difference between the two mass 
definitions is a perturbative series multiplying the scale 
$\mu_c^{\rm PS}=\epsilon^{\rm PS} m_b^{\rm PS}$. At NLO in $\alpha_s$, 
it then follows from (\ref{C0rela}) that
\begin{eqnarray}
   C_0^{\rm PS}(\epsilon^{\rm PS}) &=& C_0^{\rm pole}(\epsilon^{\rm PS})
    \left[ 1 + \frac{4\alpha_s(\mu)}{3\pi}\,
    \frac{\mu_c^{\rm PS}}{\epsilon^{\rm PS} m_b^{\rm PS}}\,
    \frac{(1-2\epsilon^{\rm PS})(3-4\epsilon^{\rm PS})}
         {1-\epsilon^{\rm PS}} \right] \nonumber\\
   &=& C_0^{\rm pole}(\epsilon^{\rm PS})
    \left[ 1 + \frac{4\alpha_s(\mu)}{\pi} + O(\epsilon^{\rm PS})
    \right] \,.
\end{eqnarray}
From now on we will use the PS mass 
$m_b\equiv m_b^{\rm PS}(\mu_c^{\rm PS})$ in all our equations and omit 
the label ``PS'' on the quantities $m_b$, $\mu_c$ and $\epsilon$. Using 
(\ref{C0pole}) and adding the extra contribution proportional to 
$\alpha_s(\mu)$, we obtain
\begin{equation}\label{C0PS}
   C_0^{\rm PS}(\epsilon)
   \simeq \left( \frac{\alpha_s(\mu)}{\alpha_s(m_b)} \right)^\frac{12}{25}
   \left[ 1 - 0.71\,\frac{\alpha_s(m_b)}{\pi} 
   + \frac{\alpha_s(\mu)}{\pi} \left( 0.63
   + 2\ln\frac{\mu}{\mu_c} \right) \right] + O(\epsilon) \,.
\end{equation}
The introduction of the PS mass has much reduced the size of the NLO 
correction. The result for the leading contribution to $C_0(\epsilon)$ 
in the PS scheme is shown by the solid line in Figure~\ref{fig:4}. It 
exhibits a better stability that in the on-shell scheme, and it is 
stable down to lower values of the renormalization scale.

The value of the PS mass at the scale $\mu_2=2$\,GeV has been 
determined from a sum-rule analysis of the $b\bar b$ production cross
section near threshold, with the result $m_b^{\rm PS}(2\,{\rm GeV})
=(4.59\pm 0.08)$\,GeV  (corresponding to 
$\overline{m}_b(m_b)=(4.25\pm 0.08)$\,GeV in the $\overline{\rm MS}$ 
scheme) \cite{BeSi}. At NLO, we can use relation (\ref{PSdef}) to 
convert this into a value of the PS mass at the scale $\mu_c$. This 
gives the implicit equation
\begin{eqnarray}
   m_b^{\rm PS}(\mu_c) &=& m_b^{\rm PS}(\mu_2) 
    + \mu_c\,\frac{4\alpha_s(\mu_2)}{3\pi} \left[ 
    \left( \frac{\mu_2}{\mu_c} - 1 \right) \left( 1 
    + \frac{203}{36}\,\frac{\alpha_s(\mu_2)}{\pi} \right)
    - \frac{25}{6}\,\frac{\alpha_s(\mu_2)}{\pi}
    \ln\frac{\mu_2}{\mu_c} \right] \nonumber\\
   &\stackrel{!}{=}& \mu_c + \sqrt{\mu_c^2 + q_0^2} \,,
\end{eqnarray}
from which we determine the scale $\mu_c$ and then the mass 
$m_b=m_b^{\rm PS}(\mu_c)$. For example, we find $m_b\simeq 4.73$\,GeV, 
$\mu_c\simeq 1.13$\,GeV, $\epsilon\simeq 0.24$ for $q_0^2=(M_B-M_D)^2$, 
and $m_b\simeq 4.79$\,GeV, $\mu_c\simeq 0.83$\,GeV, 
$\epsilon\simeq 0.17$ for $q_0^2=15$\,GeV$^2$.

In (\ref{C0PS}) we have obtained a RG-improved expression for the 
short-distance coefficient at leading order in $\epsilon$. It is at 
present not possible to extend this analysis to higher orders in 
the hybrid expansion, since the corresponding two-loop anomalous 
dimensions of higher-dimensional operators are unknown. However, since 
in the PS scheme the leading term in $\epsilon$ gives the dominant 
contribution to $C_0(\epsilon)$, we expect that the unresolved scale 
ambiguity in the higher-order terms does not introduce a large 
uncertainty. Our final expression for the short-distance coefficient 
at NLO is
\begin{equation}\label{C0final}
   C_0^{\rm PS}(\epsilon) \simeq 
   \left( \frac{\alpha_s(\mu)}{\alpha_s(m_b)} \right)^\frac{12}{25}
   \left[ 1 - 0.71\,\frac{\alpha_s(m_b)}{\pi} 
   + \frac{\alpha_s(\mu)}{\pi} \left( 0.63 + 2\ln\frac{\mu}{\mu_c}
   \right) + \frac{\epsilon\bar\alpha_s}{\pi}\,G(\epsilon) \right] \,,
\end{equation}
where the scale in $\bar\alpha_s$ in the $O(\epsilon)$ term remains 
undetermined. The exact result for the function $G(\epsilon)$ in the
PS scheme is
\begin{eqnarray}
   G(\epsilon) &=& \frac{1}{12(1-\epsilon)\epsilon^4} \left[
    L_2(1-2\epsilon) - \frac{\pi^2}{6}
    - \left( \frac{13}{12} - 8\epsilon + 28\epsilon^2
    - \frac{128}{3}\epsilon^3 + 20\epsilon^4 \right)
    \ln(1-2\epsilon) \right] \nonumber\\
   &&\mbox{}- \frac{1}{6(1-\epsilon)\epsilon^3} \left[
    \frac{1}{12} - \frac{89}{12}\epsilon + 21\epsilon^2 + 40\epsilon^3
    - 64\epsilon^4 + \left( 1 + \epsilon + \frac43\epsilon^2
    + 2\epsilon^3 \right) \ln2\epsilon \right] \nonumber\\
   &&\mbox{}+ \frac{4}{3\epsilon}
    \left[ L_2(2\epsilon) - L_2(1-2\epsilon) + \frac{\pi^2}{6} \right]
    \nonumber\\
   &=& - \frac83\ln 2\epsilon - \frac{95}{18}
    + \epsilon \left( - \frac{32}{15}\ln 2\epsilon + \frac{1337}{450}
    \right) + O(\epsilon^2) \,.
\end{eqnarray}
The left-hand plot in Figure~\ref{fig:C0} shows $C_0(\epsilon)$ for the 
optimal choice $q_0^2=(M_B-M_D)^2$ as a function of the renormalization 
scale. The width of the band reflects the sensitivity of the result to 
the value of the coupling $\bar\alpha_s$ associated with the 
$O(\epsilon)$ terms in (\ref{C0final}), which we vary between 
$\alpha_s(m_b)$ and $\alpha_s(\mu)$. To estimate the residual scale 
dependence we vary $\mu$ between the values $\mu_c$ and $2\mu_c$. For 
lower values the perturbative expansion diverges, since the running 
coupling $\alpha_s(\mu)$ strongly increases below $\mu\approx 1$\,GeV. 
For comparison, we mention that the naive perturbative analysis with 
fixed scale $\mu=m_b$ adopted in \cite{BLL} would give the much smaller 
value $C_0(\epsilon)\simeq 1.16$ at minimal $q_0^2\simeq 11.6$\,GeV$^2$. 
The fact that we find larger QCD corrections will have important 
implications for the extraction of $|V_{ub}|$. The right-hand plot in 
Figure~\ref{fig:C0} shows $C_0(\epsilon)$ as a function of $q_0^2$. The 
width of the band represents the total scale uncertainty, estimated 
by variation of $\mu$ and $\bar\alpha_s$ as described above. 

\FIGURE[t]{\epsfig{file=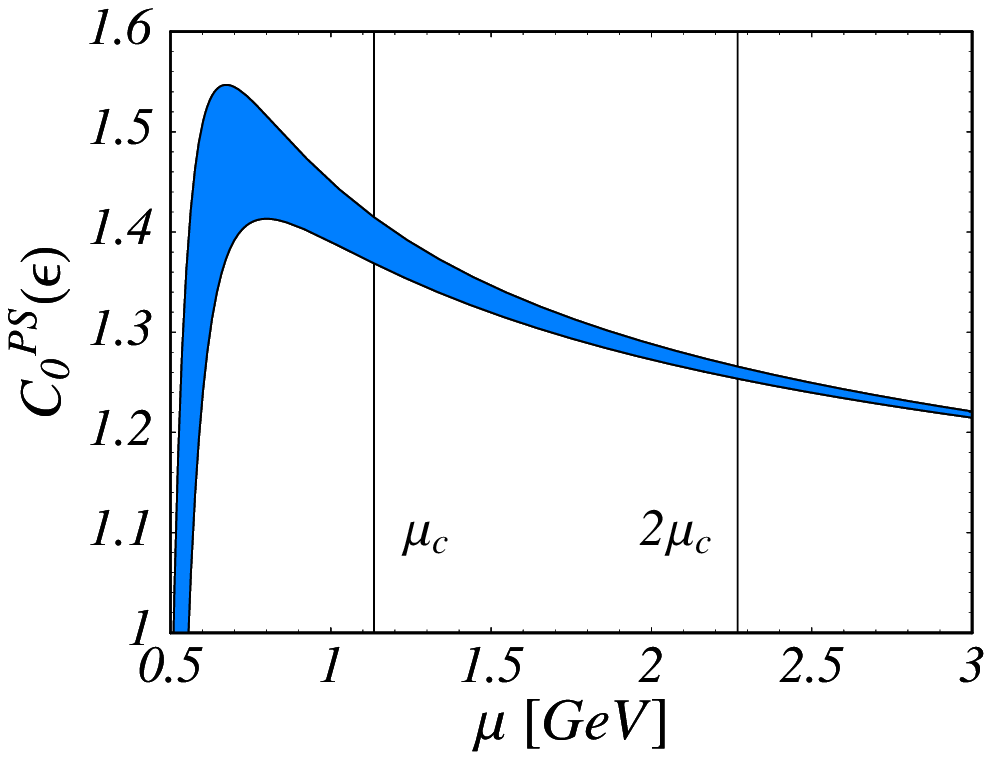,width=7.1cm}
\epsfig{file=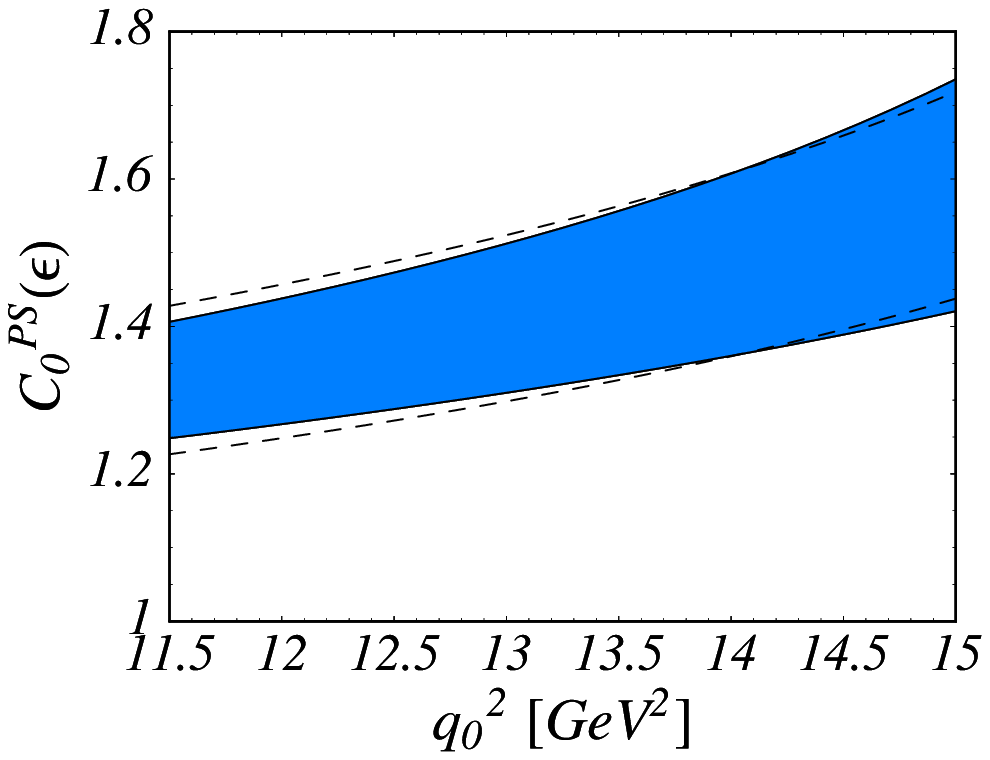,width=7.1cm}
\caption{\label{fig:C0}
Left: Scale dependence of the coefficient $C_0^{\rm PS}(\epsilon)$ for 
$q_0^2=(M_B-M_D)^2$. The width of the band indicates the scale 
ambiguity of higher-order terms in the hybrid expansion. The vertical 
lines show the window between $\mu=\mu_c$ and $\mu=2\mu_c$ used to 
estimate the scale dependence. Right: Coefficient $C_0(\epsilon)$ 
including theoretical uncertainties, as a function of the lepton 
invariant mass cut (see text).}}

An independent way to estimate the uncertainty in the value of the 
perturbative coefficient $C_0(\epsilon)$ is based on the contour 
representation (\ref{contour}) for the fraction $F(\epsilon)$. As we 
have just discussed, the value of $C_0(\epsilon)$ depends on the 
definition of the heavy-quark mass. However, the variation of the 
$O(\alpha_s)$ correction in (\ref{tfun}) along the circle in the 
complex momentum plane is independent of mass redefinitions. We may 
thus take the $\varphi$-variation of the one-loop correction, given
by $\alpha_s(\mu_c)/3\pi$ times the variation of the real part of the 
function $Y(e^{i\varphi},\epsilon)$ in (\ref{Yfun}), as a typical
size of an $O(\alpha_s)$ correction in the problem at hand. For an 
asymptotic series, the value of that correction provides an estimate 
for the magnitude of unknown higher-order corrections. The dashed 
lines in the right-hand plot in Figure~\ref{fig:C0} show this 
variation as an error band applied to the central values of 
$C_0(\epsilon)$. This independent evaluation of higher-order effects 
is in good agreement with our previous estimate of the perturbative 
uncertainty, giving us confidence that this estimate is a realistic 
one.

\section{Higher-order power corrections}

Uncertainties enter the theoretical prediction for the fraction 
$F(\epsilon)$ also at the level of power correction. First, there are
unknown $O(\alpha_s)$ corrections to the Wilson coefficient 
$C_2(\epsilon)$ multiplying the term proportional to $\lambda_2/\mu_c^2$ 
in (\ref{Feps}). To estimate their effect, we replace the bracket 
$[\dots]$ in this equation with
$C_0(\epsilon)\,[1-\frac32\,C_2(\epsilon)\,\lambda_2/\mu_c^2+\dots]$, 
which amounts to multiplying the tree-level coefficient $C_2(\epsilon)$ 
in the original expression with $C_0(\epsilon)$. At $q_0=(M_B-M_D)^2$, 
the difference is a $5\%$ effect. Potentially more important are 
higher-order power corrections scaling as $(\Lambda/\mu_c)^3$. The 
operator matrix elements contributing at third order in the heavy-quark 
expansion can be identified \cite{GrKa,Bauer}, but little is known 
about their actual size. Naive dimensional analysis suggests that, with 
a typical hadronic scale $\Lambda\approx 0.5$\,GeV, a third-order power 
correction could be of order $(\Lambda/\mu_c)^3\sim 0.09$ for 
$q_0^2=(M_B-M_D)^2$ and $(\Lambda/\mu_c)^3\sim 0.22$ for 
$q_0^2=15$\,GeV$^2$, but clearly these are rough estimates which must 
be taken with caution. 

\FIGURE[t]{\epsfig{file=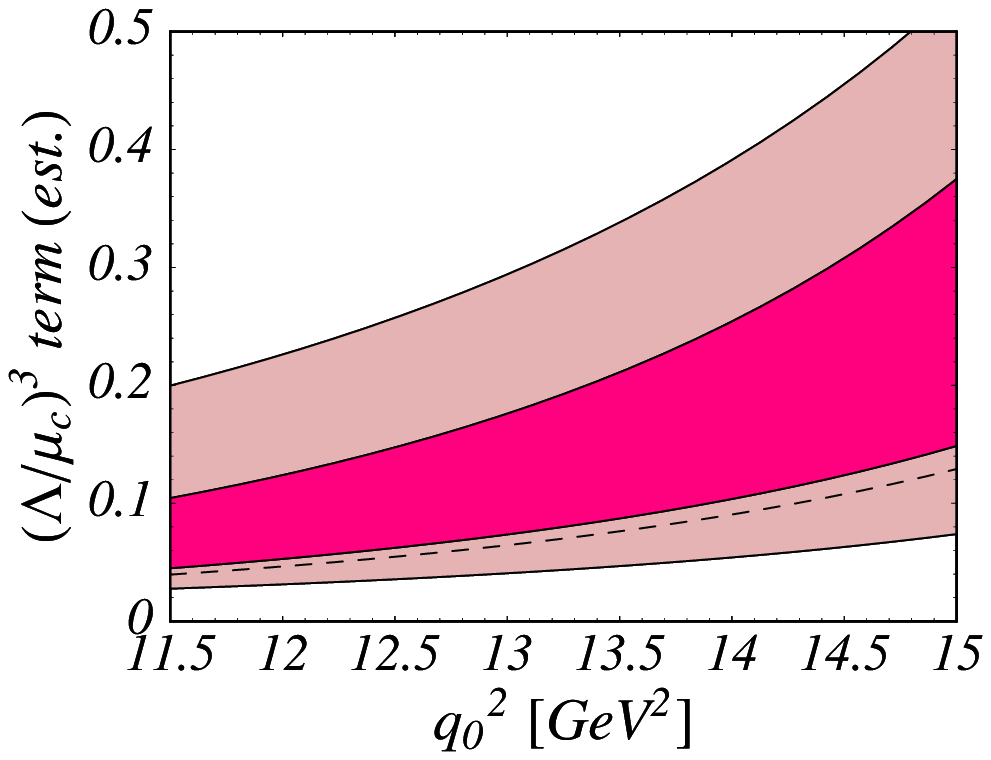,width=7.1cm}
\caption{\label{fig:power}
Estimates of third-order power corrections to the function 
$F(\epsilon)$, in units of the tree-level contribution (see text).}}

We will attempt to extract as much information about power corrections 
as possible from the formulae derived in Section~\ref{sec:tensor} for 
the $B\to X\,l\,\nu$ decay rate and spectra in the presence of a lepton
invariant mass cut. We start with the quantity $F(\epsilon)$ itself, 
which as shown in (\ref{Feps}) receives a moderate second-order power 
correction proportional to the parameter $\lambda_2\simeq 0.12$\,GeV$^2$. 
The dashed line in Figure~\ref{fig:power} shows an estimate of the 
unknown $(\Lambda/\mu_c)^3$ correction, obtained by raising this 
second-order term to the power 3/2. To address the question to what 
extent this is a conservative estimate of a ``generic'' higher-order 
correction, we focus on the differential spectrum in (\ref{xirate}) and 
on the contour representation in (\ref{contour}). The function
$F(\epsilon)$ is obtained from these results by performing integrals 
over $\xi$ or over the contour in the complex plane, respectively. 
However, the differential distributions contain additional information 
about power corrections, which is not seen after the integrations 
are performed. We first discuss the case of the contour integral in 
(\ref{contour}), taking the point of view that except for the region of 
small $\varphi$ the magnitude of $t(e^{i\varphi},\epsilon)$ can be used 
to estimate the ``generic'' size of corrections to the heavy-quark 
limit. This is so because on any point on the circle far away from the 
real, positive $\hat p^2$ axis, the function $t(e^{i\varphi},\epsilon)$ 
admits an operator product expansion in a series of local operators, 
whose contributions scale like powers of $\Lambda/\mu_c$. The average 
over the circle determines the corrections to the function $F(q_0^2)$. 
However, the finer details of the distribution on the circle become 
relevant, e.g., when in a real experiment events with different 
hadronic masses and energies are weighted by different efficiencies. In 
other words, the various terms proportional to $\lambda_1/\mu_c^2$ and 
$\lambda_2/\mu_c^2$ in (\ref{tfun}) are as valid as estimate of a 
second-order power correction as is the $\lambda_2/\mu_c^2$ term in 
(\ref{Feps}). Specifically, we calculate the average value of the 
modulus of the power corrections in (\ref{tfun}) on the circle in the 
complex plane, and then we raise this number to the power 3/2 to obtain 
an estimate of a ``generic'' $(\Lambda/\mu_c)^3$ correction. For our 
numerical analysis we use $\lambda_1=-(0.30\pm 0.15)$\,GeV$^2$, which 
is in the ball park of recent determinations \cite{Braun}. The result 
is shown by the dark band in Figure~\ref{fig:power}, whose width 
reflects the sensitivity to the value of $\lambda_1$. If we were to 
consider larger values of $|\lambda_1|$ the upper limit of the band 
would increase. Another estimate of power corrections can be obtained 
from the coefficients of the various $\delta$-function terms in 
(\ref{xirate}). If we take one third of the geometric average of the 
three coefficients, and raise the result to the power 3/2, we obtain 
the light band shown in the figure.

The above analysis shows that, as expected, there is a large 
uncertainty in the estimate of higher-order power corrections. We do 
not claim that these corrections are likely to be as large as indicated 
by the upper limit of the light band in Figure~\ref{fig:power}, 
however, as we have shown this would indeed be possible without 
introducing any unnaturally large coefficients or parameters. Keeping 
this caveat in mind, we will from now on use the upper limit of the 
dark band in the figure as our estimate of third-order power 
corrections. Numerically, this estimate is close to the one obtained 
by BLL \cite{BLL}.

\section{Phenomenological implications and summary}

The proposal of BLL is to use the theoretical calculation of the
fraction $F(q_0^2)$ to obtain a model-independent determination of the 
CKM matrix element $|V_{ub}|$ with controlled and small theoretical 
uncertainty \cite{BLL}. To this end, one uses the relation 
${\rm Br}(B\to X_u\,l\,\nu)|_{q^2>q_0^2}
=F(q_0^2)\,\Gamma(B\to X_u\,l\,\nu)\,\tau_B$, where $\tau_B$ is the 
$B$-meson lifetime, and $\Gamma(B\to X_u\,l\,\nu)$ is the total 
semileptonic decay rate into charmless final states. This rate can be 
calculated with high accuracy in terms of a low-scale subtracted 
$b$-quark mass, including perturbative corrections of order 
$[\alpha_s(m_b)]^2$ \cite{Ritb} and power corrections of order 
$(\Lambda/m_b)^2$. For our purposes, we use the PS mass defined at the 
scale $\mu_2=2$\,GeV. Then the expression for the total rate is 
\cite{Martin}
\begin{equation}
   \Gamma(B\to X_u\,l\,\nu)
   = \frac{G_F^2 |V_{ub}|^2 [m_b^{\rm PS}(\mu_2)]^5}{192\pi^3}\,
   (1 + \delta_{\rm pert} + \delta_{\rm power}) \,,
\end{equation}
where $\delta_{\rm pert}\simeq 0.04$ at two-loop order, and 
$\delta_{\rm power}=\frac{\lambda_1-9\lambda_2}{2m_b^2}\simeq-0.03$. 
The small uncertainties in these two quantities are negligible for
our numerical analysis below. Using these results, we obtain the master 
formula
\begin{equation}\label{master}
   |V_{ub}| = 2.96\times 10^{-3} \left[
   \frac{{\rm Br}(B\to X\,l\,\nu)|_{q^2>q_0^2}}{10^{-3}\,F'(q_0^2)}\,
   \frac{1.6\,{\rm ps}}{\tau_B} \right]^{1/2} ,
\end{equation}
where all theoretical uncertainties are contained in the function
\begin{equation}
   F'(q_0^2) = \left( \frac{m_b^{\rm PS}(\mu_2)}{4.59\,{\rm GeV}}
   \right)^5 F(q_0^2) \,.
\end{equation}
The mass dependence due to factor $[m_b^{\rm PS}(\mu_2)]^5$ from the 
total decay rate is positively correlated with the mass dependence of 
the function $F(q_0^2)$. As a result, our predictions for the function 
$F'(q_0^2)$ become extremely sensitive to the value of the $b$-quark 
mass. For practical purposes, this dependence can be parameterized as
\begin{equation}
   F'(q_0^2) \propto \left( \frac{m_b^{\rm PS}(\mu_2)}{4.59\,{\rm GeV}}
   \right)^{\Delta(q_0^2)} ,
\end{equation}
where
\begin{equation}
   \Delta(q_0^2) \simeq 10
   + \frac{q_0^2-(M_B-M_D)^2}{1\,\mbox{GeV}^2} \,.
\end{equation}
In Table~\ref{tab:2}, we show our final results for the quantity 
$F'(q_0^2)$ and its theoretical uncertainties (as estimated above) for 
some representative values of $q_0^2$. For comparison, we note that 
BLL obtained the values $F'((M_B-M_D)^2)=0.169\pm 0.016$ and 
$F'(15\,{\rm GeV}^2)=0.061\pm 0.013$, where the dominant theoretical
error was assumed to be due to higher-order power corrections. Our 
central values are significantly higher because of the larger 
perturbative correction obtained after RG improvement. Note that our 
error estimates are about 2.5 times as large as those quoted by BLL. 
The difference between the central values of the two calculations is 
about $1\sigma$ of our errors, and about $2\sigma$ of their errors. 

\TABULAR[t]{|c|c|cccc|}
{\hline
$\phantom{\bigg|} q_0^2 \phantom{\bigg|}$ & $F'(q_0^2)$ & $\delta m_b$
 & $\alpha_s^2$ & $\alpha_s (\Lambda/\mu_c)^2$
 & $(\Lambda/\mu_c)^3$ \\
\hline
$\phantom{\bigg|} (M_B-M_D)^2 \phantom{\bigg|}$ & $0.204\pm 0.040$
 & 16.7\% &  6.0\% & 3.0\% & 8.2\% \\[0.1cm]
13\,GeV$^2$   & $0.151\pm 0.036$ & 18.5\% &  7.2\% & 4.7\% & 12.5\%
 \\[0.15cm]
15\,GeV$^2$   & $0.090\pm 0.032$ & 22.2\% & 10.0\% & 9.4\% & 23.8\%
 \\[0.1cm]
\hline}
{\label{tab:2}
Predictions for $F'(q_0^2)$ for three values of the lepton invariant 
mass cut. The right-hand portion contains the relative theoretical 
uncertainties due to the uncertainty in the value of the $b$-quark 
mass (assuming $\delta m_b=80$\,MeV \protect\cite{BeSi}), 
higher-order perturbative corrections (including scale dependence), 
perturbative corrections to second-order power corrections, and 
higher-order power corrections.}

In Figure~\ref{fig:final} we show a graphical representation of the
fraction $F'(q_0^2)$ and its total theoretical uncertainty. This 
result, together with the master formula (\ref{master}), provides the
theoretical basis for the determination of $|V_{ub}|$. The right-hand
plot in the figure shows the fractional theoretical uncertainty in the 
result for $|V_{ub}|$. Although our error estimates are more pessimistic
than those presented by BLL, we still conclude that their method 
provides a very promising route for a precise determination of 
$|V_{ub}|$. For a realistic cut on the lepton invariant mass in the
vicinity of $q_0^2\simeq 12.5$\,GeV$^2$, which is about 1\,GeV$^2$ 
above the optimal value, the theoretical uncertainty in $|V_{ub}|$ is 
close to 10\%. A determination with such an accuracy would be a 
significant improvement with respect to the present knowledge of this 
important parameter. We believe it would also be more reliable than a 
future determination obtained by combining the partial decay rates in 
the endpoint regions of $B\to X_s\gamma$ and $B\to X_u\,l\,\nu$ decay
spectra \cite{shape2,Korc,Ira,Leib}, which is limited by uncontrollable 
power corrections of first order in $\Lambda/m_b$ that violate the 
factorization of soft and collinear singularities.

\FIGURE[t]{\epsfig{file=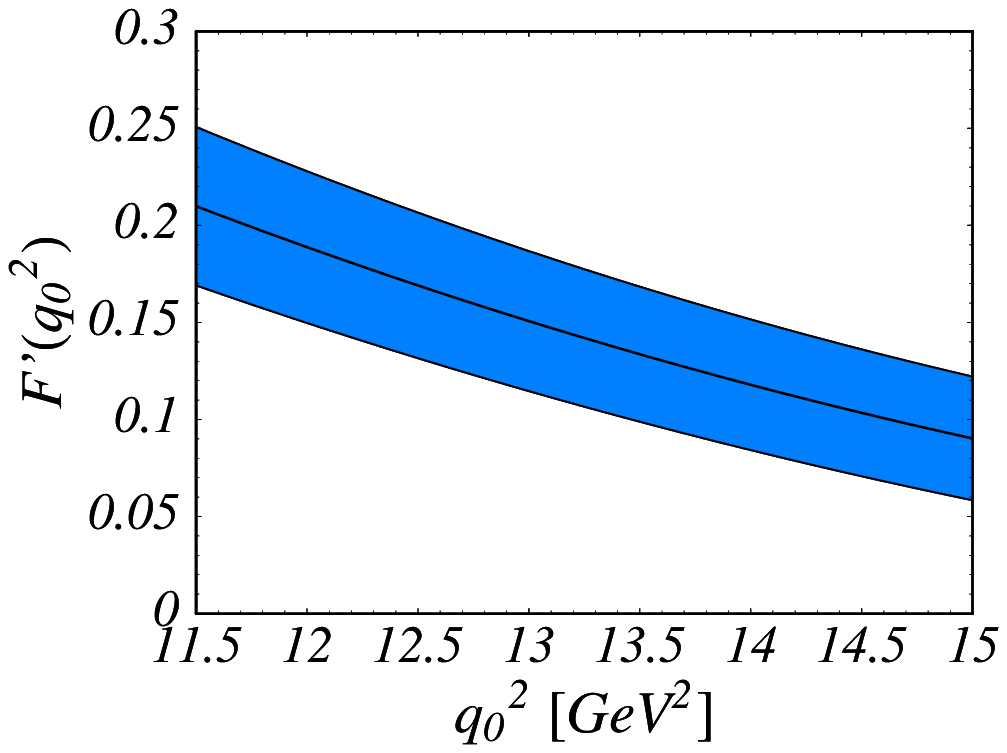,width=7.1cm}
\epsfig{file=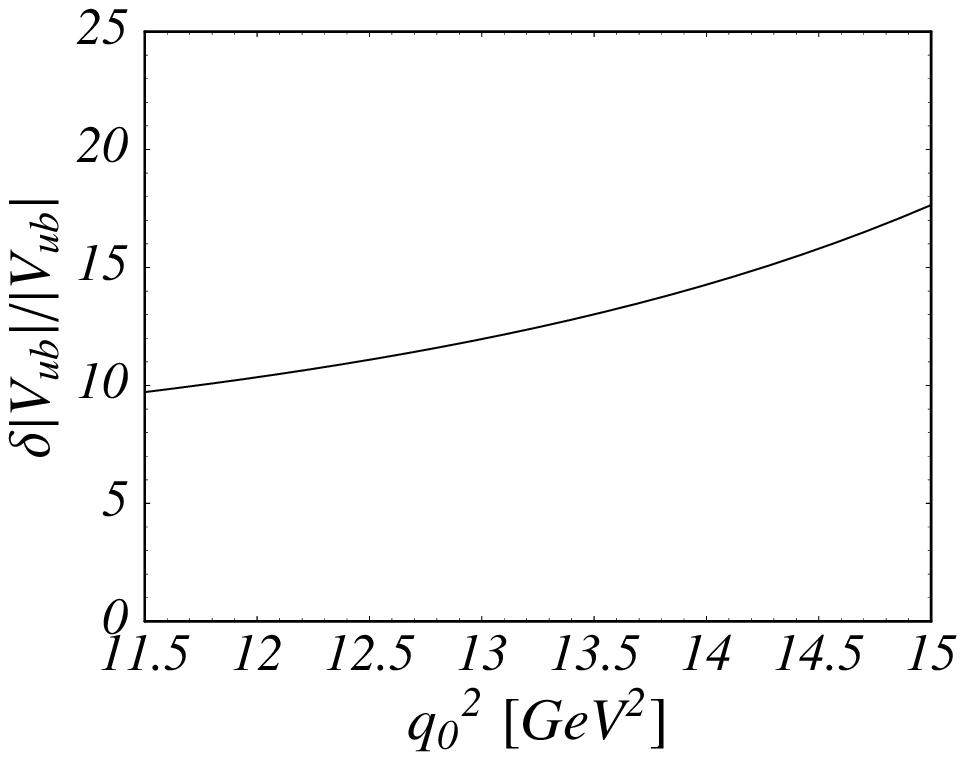,width=7.1cm}
\caption{\label{fig:final}
Left: Theoretical prediction for the quantity $F'(q_0^2)$. Right: 
Relative theoretical uncertainty in the extraction of $|V_{ub}|$.}}

According to Table~\ref{tab:2}, the dominant sources of theoretical 
uncertainty in the extraction of $|V_{ub}|$ are associated with the 
sensitivity to the value of the $b$-quark mass and with unknown, 
higher-order power corrections. Whereas it is not obvious how one
should obtain a reliable value for the power corrections, the precision 
in the value of the $b$-quark mass can presumably be improved by 
reducing the theoretical uncertainties in the analysis of $(b\bar b)$ 
bound states.\footnote{We stress, however, that using the so-called 
Upsilon mass {\em defined\/} as one half of the mass of the 
$\Upsilon(1S)$ bottomonium state \protect\cite{1Smass} does not 
eliminate the uncertainty associated with the variation of the 
$b$-quark mass. As discussed in \cite{Martin}, this choice obscures the 
presence of an unknown nonperturbative contribution to the bound-state 
mass, which is neglected in the perturbative expression of the 
$B$-meson decay rate in terms of the Upsilon mass. In other words, in 
such a scheme the value of $m_b$ is known (by definition) with very 
high precision, but for consistency the uncertainty shown in the third 
column in Table~\ref{tab:2} must then be added to the other theoretical 
uncertainties.}
In addition, it would be possible to reduce the perturbative 
uncertainty in the calculation in two ways, by calculating the exact 
$O(\alpha_s^2)$ corrections to the fraction $F(q_0^2)$ (the two-loop 
corrections to the total decay rate are known \cite{Ritb}), and by 
computing the two-loop anomalous dimensions of the operators 
contributing at $O(\epsilon)$ in the hybrid expansion. Both 
calculations are technically feasible and should be done. 

In summary, we have analyzed the structure of the heavy-quark expansion 
for the inclusive, semileptonic $B\to X\,l\,\nu$ decay rate with a 
lepton invariant mass cut $q^2\ge q_0^2$. This expansion is 
characterized by a hadronic scale $\mu_c\lsim m_c$ determined by the 
value of $q_0^2$. Because $m_b\gg\mu_c\gg\Lambda$, the heavy-quark 
expansion can be organized as a combined (hybrid) expansion in two 
small mass ratios. The physics associated with the two large scales 
$m_b$ and $\mu_c$ is disentangled using the HQET, whereas the physics 
on the scale $\mu_c$ can be separated from long-distance physics 
associated with $\Lambda$ utilizing an operator product expansion. We 
have used this formalism to obtain a RG-improved expression for the 
leading short-distance coefficient in the heavy-quark expansion at NLO. 
The summation of large logarithms in the hybrid expansion turns out to 
be important and strongly enhances the overall size of the perturbative 
correction. We have also emphasized that in order to obtain a stable 
perturbative prediction it is important to eliminate the $b$-quark pole 
mass in favor of a low-scale subtracted quark mass, such as the PS mass. 
Finally, we have presented several independent estimates of higher-order 
power corrections in the heavy-quark expansion, which at present do not 
permit a rigorous treatment. We find that with realistic values of the
lepton invariant mass cut the overall theoretical uncertainty in the 
extraction of $|V_{ub}|$ is about 10\%, which is larger than previously 
estimated but still significantly less than the current uncertainty in 
this parameter. 

\acknowledgments
I am grateful to Martin Beneke, Alex Kagan, Zoltan Ligeti and Mark Wise 
for useful discussions. This work was supported in part by the National 
Science Foundation.

\end{document}